\documentclass[manuscript]{aastex}

\shorttitle{Background Quasars in the Vicinity of M\,31 and M\,33}
\shortauthors{Huo et al.}

\begin{document}

\title{The LAMOST Survey of Background Quasars in the Vicinity of the Andromeda and
Triangulum Galaxies -- II. Results from the Commissioning Observations and the Pilot Surveys}

\author{
Zhi-Ying Huo\altaffilmark{1},         Xiao-Wei Liu\altaffilmark{2,3},         Mao-Sheng Xiang\altaffilmark{3},  
Hai-Bo Yuan\altaffilmark{2,8},       Yang Huang\altaffilmark{3},            Hui-Hua Zhang\altaffilmark{3},           
Lin Yan\altaffilmark{4},                 Zhong-Rui Ba\altaffilmark{1},         Jian-Jun Chen\altaffilmark{1},
Xiao-Yan Chen\altaffilmark{1},      Jia-Ru Chu\altaffilmark{5},             Yao-Quan Chu\altaffilmark{5},
Xiang-Qun Cui\altaffilmark{6},       Bing Du\altaffilmark{1},                  Yong-Hui Hou\altaffilmark{6},
Hong-Zhuan Hu\altaffilmark{5},    Zhong-Wen Hu\altaffilmark{6},       Lei Jia\altaffilmark{1},
Fang-Hua Jiang\altaffilmark{6},    Ya-Juan Lei\altaffilmark{1},            Ai-Hua Li\altaffilmark{6}, 
Guang-Wei Li\altaffilmark{1},        Guo-Ping Li\altaffilmark{6},            Jian Li\altaffilmark{1}, 
Xin-Nan Li\altaffilmark{6},             Yan Li\altaffilmark{7},                     Ye-Ping Li\altaffilmark{6}, 
Gen-Rong Liu\altaffilmark{6},       Zhi-Gang Liu\altaffilmark{5},          Qi-Shuai Lu\altaffilmark{6}, 
A-Li Luo\altaffilmark{1},                Yu Luo\altaffilmark{1},                    Li Men\altaffilmark{1}, 
Ji-Jun Ni\altaffilmark{6},                Yong-Jun Qi\altaffilmark{6},           Zhao-Xiang Qi\altaffilmark{7}, 
Jian-Rong Shi\altaffilmark{1},        Huo-Ming Shi\altaffilmark{1},         Shi-Wei Sun\altaffilmark{1}, 
Zheng-Hong Tang\altaffilmark{7}, Yuan Tian\altaffilmark{1},               Liang-Ping Tu\altaffilmark{1}, 
Dan Wang\altaffilmark{1},              Feng-Fei Wang\altaffilmark{1},      Gang Wang\altaffilmark{1}, 
Jia-Ning Wang\altaffilmark{6},       Lei Wang\altaffilmark{6},                Shu-Qing Wang\altaffilmark{1}, 
You Wang\altaffilmark{6},              Yue-Fei Wang\altaffilmark{6},        Ming-Zhi Wei\altaffilmark{1}, 
Yue Wu\altaffilmark{1},                  Xiang-Xiang Xue\altaffilmark{1},       Zheng-Qiu Yao\altaffilmark{6}, 
Yong Yu\altaffilmark{7},                 Hui Yuan\altaffilmark{1},                Chao Zhai\altaffilmark{5}, 
En-Peng Zhang\altaffilmark{1},       Hao-Tong Zhang\altaffilmark{1},  Jian-Nan Zhang\altaffilmark{1}, 
Wei Zhang\altaffilmark{1},               Yan-Xia Zhang\altaffilmark{1},     Yong Zhang\altaffilmark{6}, 
Zhen-Chao Zhang\altaffilmark{6},   Gang Zhao\altaffilmark{1},           Ming Zhao\altaffilmark{7}, 
Yong-Heng Zhao\altaffilmark{1},     Fang Zhou\altaffilmark{6},           Xin-Lin Zhou\altaffilmark{1}, 
Yong-Tian Zhu\altaffilmark{6},        Si-Cheng Zou\altaffilmark{1}
}

\altaffiltext{1}{Key Laboratory of Optical Astronomy, National Astronomical Observatories, Chinese Academy of Sciences,
             Beijing 100012, P.R. China; Email: zhiyinghuo@bao.ac.cn}
\altaffiltext{2}{Kavli Institute for Astronomy and Astrophysics, Peking University,
             Beijing 100871, P.R. China}
\altaffiltext{3}{Department of Astronomy, Peking University,
              Beijing 100871, P.R. China}
\altaffiltext{4}{Infrared Processing and Analysis Center, California Institute of Technology, MS 100-22, Pasadena, CA 91125, USA}              
\altaffiltext{5}{University of Science and Technology of China, Hefei 230026, P.R. China}
\altaffiltext{6}{Nanjing Institute of Astronomical Optics \& Technology, National Astronomical Observatories, 
             Chinese Academy of Sciences, Nanjing 210042, P.R. China}     
\altaffiltext{7}{Shanghai Astronomical Observatory, Chinese Academy of Sciences, Shanghai 200030, P.R. China}                   
\altaffiltext{8}{LAMOST Follow}

\begin{abstract}
We present new quasars discovered in the vicinity of the Andromeda
and Triangulum galaxies with the LAMOST (Large Sky Area Multi-Object Fiber
Spectroscopic Telescope, also named Guoshoujing Telescope) during the 2010 and
2011 observational seasons.  
Quasar candidates are selected based on the available Sloan Digital Sky Survey (SDSS),
Kitt Peak National Observatory (KPNO) 4 m telescope, 
Xuyi Schmidt Telescope Photometric Survey (XSTPS) optical, 
and Wide-field Infrared Survey Explorer (WISE) near infrared photometric data. 
We present 509 new quasars discovered in a
stripe of $\sim$135 deg$^2$ from M\,31 to M\,33 along the Giant Stellar Stream
in the 2011 pilot survey datasets, and also 17 new quasars discovered in an area of $\sim$ 100
deg$^2$ that covers the central region and the southeastern halo of M31 in the
2010 commissioning datasets.  These 526 new quasars have $i$ magnitudes ranging
from 15.5 to 20.0, redshifts from 0.1 to 3.2. They represent a
significant increase of the number of identified quasars in the vicinity of M\,31 and
M\,33.
There are now 26, 62 and 139 known quasars in this region of the sky 
with $i$ magnitudes brighter than
17.0, 17.5 and 18.0 respectively, of which 5, 20 and 75 are newly-discovered. 
These bright quasars provide an invaluable collection with which to probe the kinematics and 
chemistry of the interstellar/intergalactic medium (ISM/IGM) in the Local Group of galaxies.
A total of 93 quasars are now known with locations within 2.5$^\circ$
of M\,31, of which 73 are newly discovered. 
Tens of quasars are now known to be located behind the Giant Stellar Stream,
and hundreds behind the extended halo and its associated substructures of
M\,31.  The much enlarged sample of known quasars in the vicinity of M\,31 and M\,33 can
potentially be utilized to construct a perfect astrometric reference frame to
measure the minute proper motions (PMs) of M\,31 and M\,33, along with the PMs
of substructures associated with the Local Group of galaxies.  Those PMs
are some of the most fundamental properties of the Local Group.  
\end{abstract}


\section{Introduction}


Being the most luminous member of the Local Group and the nearest archetypical
spiral galaxy, M\,31 serves as the best astrophysical laboratory for the
studies of the physical and astrophysical processes that govern the structure,
kinematics and the formation and evolution of distant galaxies.  Recent deep
optical surveys have revealed complex substructures within hundreds of
kiloparsec (kpc) of M\,31, with some of them stretching from M\,31 all the way
to M\,33, which is about 200 kpc from M\,31, suggesting a possible recent close
encounter of the two galaxies (Ibata et al. 2007; McConnachie et al.\ 2009).
Detailed chemical and kinematic investigations of M\,31 and associated
substructures are vital for our understanding of  M\,31, and also for the theory of
galaxy formation and evolution in general.

Finding background quasars in the vicinity of M\,31 have the following two
potential important applications.  Firstly, the background quasars can serve as
an ideal reference frame which would allow highly accurate
astrometric measurement of the minute proper motion (PM) of M\,31 and
also the PMs of its associated coherent substructures. 
Secondly, absorption-line spectroscopy of bright
background quasars can be used to probe the distribution, 
chemical composition and kinematics of the interstellar medium (ISM) of M\,31, the Milky
Way and the intergalactic medium (IGM) of the Local Group of galaxies.  This
technique can be used to probe structures further out from the center of the galaxy
than would be possible with traditional stellar spectroscopy, or using H~{\sc i} 21cm
observations. In addition, M\,31 has alway been a focus of interest for astronomical
observations. Extensive photometric data spanning tens of years exist.
Therefore, identifying background quasars in the vicinity of M\,31 should also
enable the construction of a valuable sample for studying quasar variability.

The PM of M\,31 represents one of the most fundamental quantities of the Local
Group, and its accurate measurement is vital to the study of the
formation, evolution and dynamics of the Local Group, in particular for
quantifying the dark matter content.  M\,31 is known to be located at a
distance of $\sim$ 785 kpc (McConnachie et al. 2005), and is moving towards the
Milky Way with a velocity of 117 km s$^{-1}$ (Binney \& Tremaine, 1987, p.605).
However, its transverse velocity has remained unknown for a long time.  The Very Long
Baseline Array, with a PM measurement uncertainty of the order of one
micro-arcsec, has measured the distance and PM of M\,33 by water maser observations
(Brunthaler et al. 2005). However, water maser sources in M\,31 have been
discovered only recently (Darling 2011), thus we do not yet have a long enough time
baseline for a meaningful PM measurements.  Based on kinematic information
about M\,31's satellite galaxies, M\,33 and IC\,10, Loeb et al.\ (2005) and van der
Marcel \& Guhathakurta (2008) deduced theoretical estimates for the PM
of M\,31 of about 80 km s$^{-1}$ (20 $\mu$as yr$^{-1}$).  Sohn, Anderson \& van der Marel
(2012) presented the first direct PM measurements for three M\,31 fields 
($\sim$ 2.7$\times$2.7 sq.\ arcmin.\ for each field) using {\it HST} observations that
span a time baseline of 5--7 yrs. Due to the sparseness of known
background quasars, their analysis had to resort to using compact background galaxies as
reference sources, even though it was hard to obtain highly accurate positions
for those galaxies given their nature as extended sources.  Being
essentially point sources with zero PMs, background quasars can serve as perfect
reference sources for highly accurate PM measurements, provided that enough of them
with sufficient space density can be identified.

At the distance of M\,31, even a luminous red giant branch star has an $I$
magnitude fainter than 20, hence high resolution spectroscopic determinations of
their chemical composition are no easy tasks even for a 10\,m class telescope
(Ibata et al.\ 2005; Gilbert et al.\ 2009).  Absorption-line spectroscopy of
background quasars allow one to probe the distribution, chemical composition and 
kinematics of the ISM associated with M\,31, and this technique can be used to
probe material further out from the galaxy center than is possible with
traditional stellar spectroscopy or with H~{\sc i} 21cm observations
(Chemin, Carignan \& Foster 2009).  
A number of papers have been published (see Savage et al.\ 2000; Schneider et al.\ 1993 
and references therein), reporting Milky Way absorption line systems which were 
detected in quasar spectra. Analyses of those
systems have revealed great complexity of the Milky Way halo gas, which exhibits
a wide range of ionization states, chemical compositions and kinematics. Similar
absorption-line spectroscopic analysis of M\,31 background quasars will help
understand the formation history of M\,31 and the Local Group.

A number of efforts have been taken to search for background quasars behind
nearby galaxies, with the intended purpose being to study their PMs or ISM/IGM.  
Tinney, Da Costa \& Zinnecker (1997) and Tinney (1999) obtained
a sample of quasars behind the Milky Way satelite galaxies.
Kim et al.\ (2012) published 663
background quasar candidates in the Large Magellanic Cloud (LMC), Koz{\l}owski
et al.\ (2012; see also references therein) increased the number of known
background quasars to 200 behind the southern edge of the LMC. 
For M\,31, Crampton et al.\ (1997) presented several quasar candidates.  
Untill recently, Sloan Digital Sky Survey (SDSS; York et al.\ 2000) obtained 75 low-redshift quasars in 
three spectroscopic plates in two fields in the outer halo of M\,31
(c.f.  Adelman-McCarthy et al.\ 2006, 2007). 
As far as we are aware, since 2007 no further observations targeting quasars
around M\,31 have been made by SDSS.  In the first
of this series of papers we presented 14 new quasars discovered with the LAMOST
during its early commissioning phase (see Huo et al.\ 2010, hereafter Paper I).

LAMOST is a quasi-meridian reflecting Schmidt telescope with an effective aperture
of about 4m and a field of view of 5\,deg in diameter (Wang et al.\ 1996; Su et al.\ 1998; 
Xing et al.\ 1998; Zhao 2000; Cui et al.\ 2004; Zhu et al.\  2006; Cui et al.\ 2010, and 
Cui, Zhao \& Chu et al.\ 2012, c.f.\ http://www.lamost.org/website/en/).
Recording 4000 celestial object spectra simultaneously with a parallel fully controllable
fiber positioning system, LAMOST is the telescope with the highest rate of spectral 
acquisition at the present time.
After a period of two years of commissioning for the purpose of technical fine-tuning,
performance characterization and optimization, as well as scientific capability
demonstration, the LAMOST pilot surveys were initiated in October 2011 and
completed in spring the following year (see Zhao et al.\ 2012).  In this paper,
we report newly-discovered background quasars in the vicinity of M\,31 and
M\,33, based on LAMOST data obtained during the commissioning phase of 2010 as
well as during the pilot surveys from autumn 2011 to spring 2012.

\section{Candidate Selection}
\label{sect:QSOcand}

With the aim of searching for background quasars in the vicinity of M\,31
and M\,33, we selected quasar candidates based on the available optical, near
infrared (IR) photometric data in the extended area of M\,31 and M\,33. Only
point-source quasar candidates were selected in order to provide a clean sample
for the PM measurements and ISM/IGM studies.

\begin{figure}
\includegraphics[width=14.0cm,angle=0]{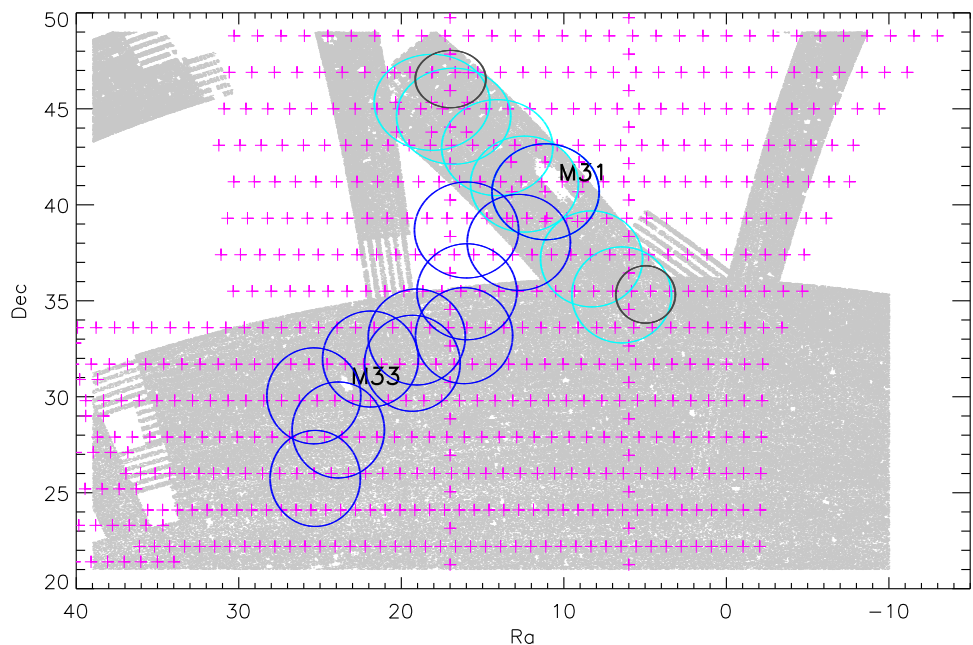}
\begin{center}
\caption{Fields observed with the LAMOST in the vicinity of M\,31 and M\,33
during the 2010 and 2011 observational seasons. 
The positions of M\,31 and M\,33 are labeled with characters, M\,31 and M\,33 are
centered in the white areas. 
The grey area represents the SDSS footprint in this area. The magenta
plus symbols delineate the centers of the individual XSTPS fields covered the M\,31/M\,33 area,
with 2$\times$2 sq.\ degrees for each field. 
The cyan and blue circles represent the LAMOST fields observed in 2010 and 2011, respectively. 
The black circles represent the three SDSS spectroscopic fields targeting quasar candidates,
the two plates in the northeast halo have the same central positions.\label{qsocand}}
\end{center}
\end{figure}

Fig.\,\ref{qsocand} shows the fields observed with the LAMOST during 2010 and
2011 observational seasons. 
M\,31 is covered by a special scan of a total area of 100 square degrees,
5$\degr$ wide and 20$\degr$ long, along the northeast-southwest direction 
(Adelman-McCarthy et al.\ 2006, 2007).  
M\,33 is covered by the SDSS southern Galactic cap survey (Aihara et al.\ 2011).  
Following Richards et al.\ (2002), low-redshift quasar candidates 
are selected as outliers from the stellar loci on the SDSS color-color diagrams
(see also Paper I for the color-color diagrams and selection criteria).
The candidates are required to have magnitude errors less than 0.2 magnitude in $ugriz$. 
With a magnitude limit set at $i \le 20.0$,  
the low-redshift quasar candidates thus selected have a number density of 27/sq.\ deg.

As shown in Fig.\,\ref{qsocand}, there are no reliable photometric data from
the SDSS in the central area of M\,31 and M\,33 given the very high stellar
densities there.  To select candidates in those crowded fields, we have made
use of  the $UBVRI$ photometric catalog of M\,31 and M\,33 provided by the
Kitt Peak National Observatory (KPNO) 4m telescope (Massey et al.\ 2006). In
doing so, we convert the KPNO $UBVRI$ magnitudes to the SDSS $ugriz$ using the
transformations provided by Jester et al.\ (2005) for quasars of redshifts $z
\leq 2.1$.

In addition to those purely optically-selected candidates, we have selected quasar
candidates combining the optical and IR photometry. Sources from Wide-field Infrared 
Survey Explorer (WISE; Wright et al.\ 2010) with colors $W1-W2 > 0.8$ were selected
(see Fig.\ 11, 12 in Yan et al.\ 2013 for the color-color distributions of different source
classification; also Stern et al.\ 2012; Wu et al.\ 2012 and references therein) 
and cross-correlated with the optical point-source catalogs of SDSS, KPNO and the 
Xuyi Schmidt Telescope Photometric Survey (XSTPS; Liu et al., in preparation, see also
http://kiaa.pku.edu.cn/DSSGAC/overview.html).  XSTPS has a component that
extends to the M\,31/M\,33 area, as shown in Fig.\,\ref{qsocand}.  In total,
about 26 quasar candidates per sq.\ degree were selected by this technique for a
limiting magnitude $i \le 20.0$. For the sky area also covered by the SDSS, we find
that about 60\% of these WISE IR quasar candidates overlap with the
optically-selected low-redshift sample. As shown in Wu et al.\ (2012), this IR-based
technique is capable of finding quasars with redshifts up to $z < 3.5$.

On the other hand, for the central regions of M\,31 and M\,33, this IR
selection algorithm also yields too many interlopers, in particular along the
optical disk of M\,31 and M\,33. We expect that most of these sources are actually
candidates asymptotic giant branch (AGB) stars, planetary nebulae (PNe),
young stellar objects (YSO) or red stars.  These are interesting objects to
study in their own right.  Further more, finding quasars in the very central
regions of M\,31 and M\,33 will be extremely valuable for the PM measurements
and ISM spectroscopic studies. As such we have included all those candidates in
our input catalogs for LAMOST observations.  The cyan and blue circles in
Fig.\,\ref{qsocand} represent the LAMOST fields observed during the 2010 and 2011
observational seasons; the sky area covered during each year was $\sim$ 100 deg$^{\rm
2}$ and $\sim$ 135 deg$^{\rm 2}$, respectively. In 2010, only optically
selected low-redshift quasars candidates based on the SDSS colors were
selected.

\section{Observations and Data Reduction}
\label{sect:Obs}

M\,31 has been targeted by the LAMOST since the commissioning phase
initiated in 2009. Paper I presented some of the early results about background
quasars discovered by LAMOST.  Here we present results based on data collected
in the 2010 and 2011 observational seasons.  

The exposure time for the plates vary from 600 to 1800s,
repeated 2 or 3 times to allow rejection of cosmic rays and to increase the
signal-to-noise ratio (S/N). 
%
In total, 731 and 5641 unique quasar candidates were observed in 2010 and 2011, respectively. 
Some sources were targeted repeatedly. This is partly because the field of view of LAMOST
is circular, which forces adjacent fields to overlap to allow coverage of a large
contiguous sky area. In 2010, only
SDSS optically-selected low-redshift quasars candidates were selected and
observed.  In October and November of 2010, LAMOST CCDs suffered severe
electromagnetic interference and insufficient dewar cooling, and so few useful
data were collected.  See Fig.\,\ref{qsocand} for the fields targeted by the
LAMOST in both 2010 and 2011.

The LAMOST has 16 low resolution spectrographs, each accommodating 250 fibers.
From the fall of 2011, LAMOST impleted slit masks with
a width of 2/3 the 3.3\,arcsec fiber diameter, yielding a spectral resolving
power $\sim$ 1800. Before that a mixture of full (i.e.\ no slit masks) and 1/2
slit modes were used, but mostly in full slit mode, which yields a spectral
resolution $\sim$ 1000. 
The light that enter each spectrograph is split into two channels, the blue
and red channels, for which the wavelength coverages are 3700-5900 ${\rm \AA}$ and
5700-9000 ${\rm \AA}$, respectively. The spectra from each spectrograph
are imaged onto two 4096$\times$4096 CCD cameras (See Cui et al.\ 2012 for further details).
The spectra were reduced using the LAMOST 2D pipeline, 
including bias subtraction, spectra tracing and extraction, wavelength
calibration, flat-fielding and sky subtraction (see Luo, Zhang \& Zhao
2004; Luo et al. 2012).
Considering the uncertainties in the commissioning phase and the pilot surveys, 
and the relatively low S/N of most quasar candidate spectra,
for the current work of identifying quasars, the non flux-calibrated spectra were used.
Hence the photometric magnitudes serve as much better
measurements of the brightness of our targets.

\begin{figure}
\includegraphics[width=13.5cm,angle=0]{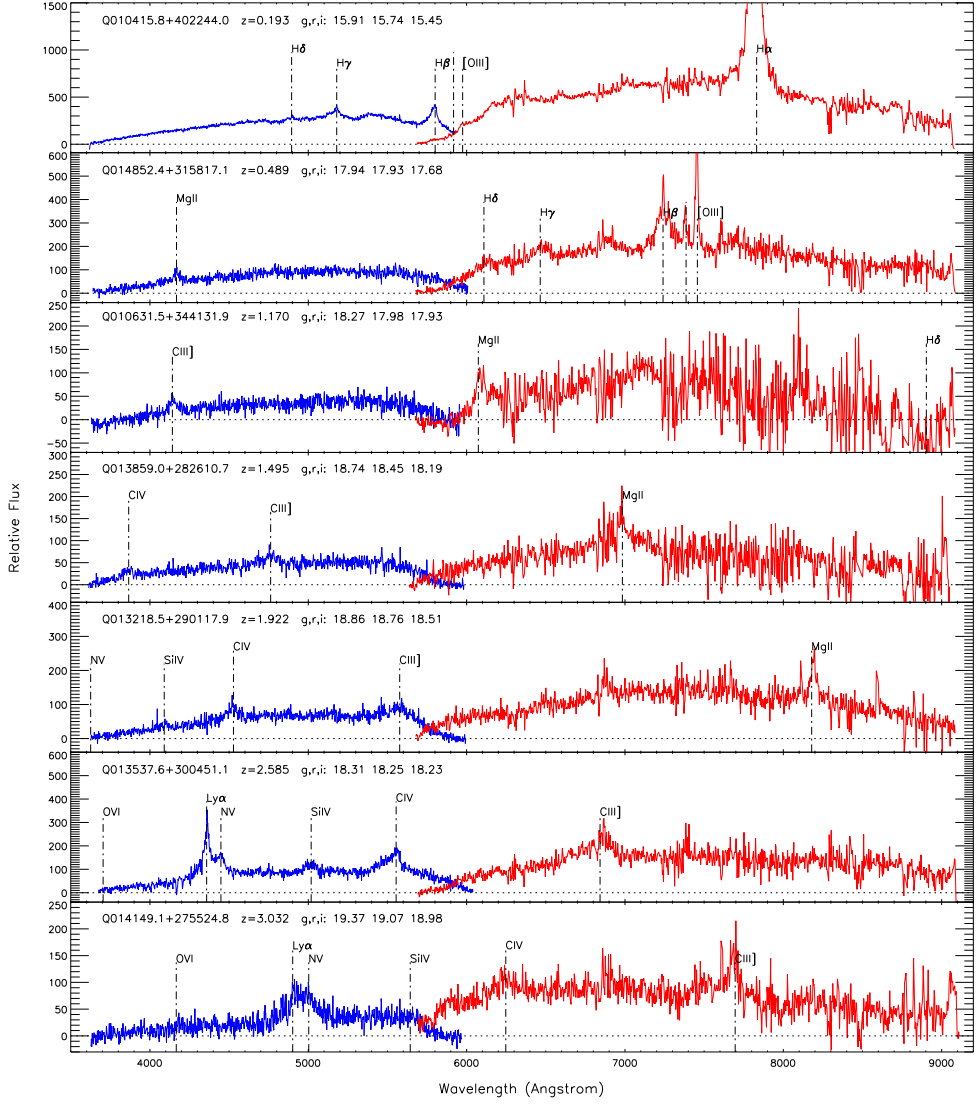}
\caption{Example of LAMOST spectra of newly-discovered quasars, with identified lines
labeled. The relative flux is in units of counts per pixel, the horizontal dotted line labeled
the zero flux. The spectra were binned by a factor of 4, and remaining cosmic rays after 
pipeline processing were removed manually for clarity.\label{spectra}}
\end{figure}

\section{Results and Discussion}
\label{sect:results}

This current work is based on observations collected during the LAMOST
commissioning phase and the pilot surveys when the LAMOST performance was
still being characterized and optimized. Thus it is unavoidable that the
data presented here suffer from some defects in one way or another, such as low
throughput, poor sky subtraction and so on. The fiber positioning,  although
improved significantly in spring 2011, still has room for further improvement.
%
Fortunately, given their characteristic broad emission line, quasars are easily identified.
We visually examined the 1D spectra of each individual quasar candidate. 
We required that at least two emission lines were
securely identified. The results led to the discovery of 17 and 509 new quasars
in the vicinity of M\,31 and M\,33 in the fields observed in 2010 and 2011,
respectively.  If we adopt a luminosity cut following the SDSS Quasar Survey
(Schneider et al.\ 2010, see also the references therein), with absolute
magnitude $M_i = -22.0$ in a cosmology $H_{\ 0} = 70$ km s$^{\rm -1}$
Mpc$^{\rm -1}$, $\Omega_{\rm M} = 0.3$, and $\Omega_{\rm \Lambda}  =
0.7$, then the number of newly-discovered quasars are 15 and 509 in 2010 and 2011 datasets,
respectively.

\begin{deluxetable}{cccccccccccccc}
\tabletypesize{\scriptsize}
\tablecaption{Catalog of New Quasars in the Vicinity of M\,31 and M\,33 Discovered by LAMOST
in Fields Observed in 2010 and 2011.\label{tab_cat}}
\tablewidth{0pt}
\tablehead{
\colhead{Object}  & \colhead{R.A.} & \colhead{Dec. (J2000)} & \colhead{Redshift} & \colhead{$u$} & \colhead{$g$} & 
\colhead{$r$} & \colhead{$i$} & \colhead{$z$} & \colhead{A(i)} & \colhead{Selection}  
} 
\startdata
  J003255.71+394619.2  &   8.232129  &  39.772018   &   1.134   &      -       &18.53   &18.24   &18.29  &     -        & 0.12 &  W,-,- \\
  J003432.52+391836.1  &   8.635497  &  39.310022   &   0.138   &  18.55   &18.53   &18.38   &17.87  & 18.03   & 0.10 &  -,S,M\\
  J003459.61+420655.2  &   8.748376  &  42.115341   &   2.472   &      -       &18.86   &18.60   &18.71  &     -        & 0.12 &  W,-,-\\
  J003506.69+404003.4  &   8.777890  &  40.667629   &   1.844   &  20.27   &20.05   &19.91   &19.56   & 19.43  & 0.14 &  -,S,-\\
  J003514.29+401414.6  &   8.809581  &  40.237413   &   0.279   &      -       &19.50   &18.74   &18.57  &     -        & 0.12 &  W,-,-\\
  J003522.68+393353.9  &   8.844538  &  39.564987   &   0.670   &      -       & 19.32  &19.48   &19.53   &    -        & 0.13 &  W,-,-\\
  J003524.49+394619.1  &   8.852052  &  39.771992   &   2.340   &      -       & 19.25  & 19.03  & 18.77  &    -        & 0.12 &  W,-,-\\
  J003535.01+404351.2  &   8.895891  &  40.730891   &   1.774   &  20.33   & 20.21  &20.25   &19.90   &20.12   & 0.13 &  -,S,-\\
  J003546.45+413358.6  &   8.943565  &  41.566297   &   2.580   &      -       & 19.21  & 18.93  & 19.02  &     -       & 0.14 &  W,-,-\\
  J003619.34+411746.5  &   9.080616  &  41.296262   &   1.651   &  20.38   & 19.90  & 19.56  &19.21   &19.09   & 0.14 &  -,S,-\\
\enddata
\tablecomments{{\it W} represents candidates selected by the IR criteria based on the WISE data,
{\it S} represents those optically selected with the SDSS criteria for low-redshift quasars, {\it M} represents
those optically selected from the KPNO 4m Local Group Galaxy survey data (Massey et al.\ 2006).
Only a portion of the Table is shown here for illustration. The whole Table
contains information about 526 new quasars, and is available online in the electronic version.}
\end{deluxetable}

All the new quasars have reasonable S/N and at least two
identified emission lines, thereby allowing reliable redshift estimates (see
Fig.\,\ref{spectra}).  In Table\,\ref{tab_cat}, we present the catalog of these
newly discovered quasars. The properties of these quasars,  including target names (in format of J{\it
hhmmss.ss+ddmmss.s}), equitorial coordinates, redshifts, the observed SDSS $u$,
$g$, $r$, $i$, $z$ magnitudes without extinction corrections, $i$-band extinction
from Schlegel et al.\ (1998), and the selection criteria, are listed.  
Only a portion of the Table is shown here. The whole Table,
containing information about all 526 new quasars, is available in the online electronic version. 
In the last column of the Table\,\ref{tab_cat}, {\it W} indicates candidates
selected with the IR criteria based on the WISE data, {\it S} indicates those
optically selected using the SDSS criteria for low-redshift quasars, and {\it
M} indicates those optically selected candidates based on the data
from the KPNO survey of Local Group Galaxies.  Columns 5-9 give the SDSS $u$, $g$, $r$,
$i$, $z$ magnitudes. For some targets only the $g$, $r$, $i$ magnitudes are
given. Those are targets selected by cross-correlating the WISE-IR candidates
with the XSTPS optical catalog.  Fig.\,\ref{spectra} shows spectra of 7 new
quasars spanning a wide range of redshifts.  The spectra have not been
flux-calibrated. At wavelengths shorter than 4000 ${\rm \AA}$, where the
instrument throughputs are low, as well as between the dichroic cross-over
wavelength range 5700--6000 ${\rm \AA}$, large spike- and trough-like artifacts
often appear in the flux-calibrated spectra. This is particularly
true for spectra with relatively low S/N, as is the case for many of our targets of interest
here.  Because of this, we have opted to not to use the flux-calibrated spectra.  Since
the LAMOST spectra were oversampled, we binned the spectra by a factor of 4 to
improve the S/N.  Cosmic rays which remained after pipeline processing were removed
manually for clarity. The catalog and the individual spectra are available at the LAMOST
public website http://www.lamost.org. 

In addition to the 526 confirmed quasars, another 20 and 193 `probable' quasars
have been identified in the current LAMOST 2010 and 2011 datasets. 
They either have marginal S/N or have only one emission line
detected and so are at uncertain redshifts. Further observations are needed to confirm
their identifications. We take this `probable' quasar catalog as a supplemental list,
this supplemental list and the related spectra are also available at the aforementioned
website, which will be updated in real time as new data are obtained
or the pipeline is improved.
Finally, 11 and 23 previously known quasars are re-observed in the 
LAMOST 2010 and 2011 datasets, respectively.

\begin{figure} 
\includegraphics[width=16cm, angle=0]{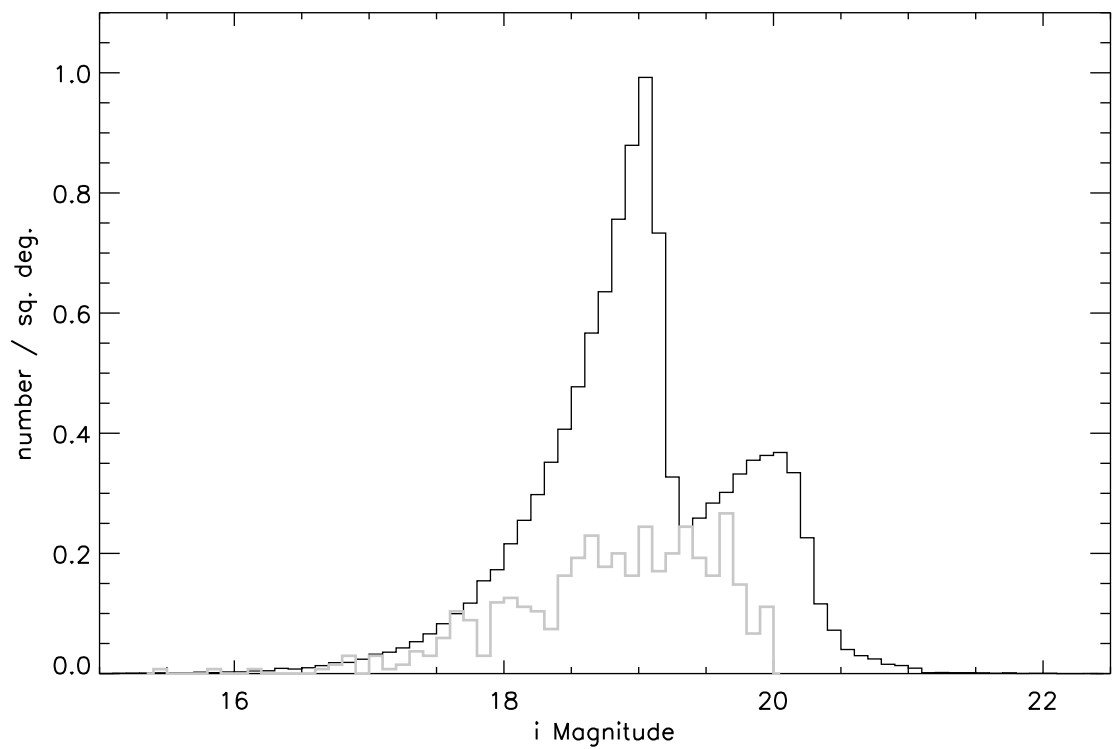}
\caption{$i$-band magnitude distributions of the 532 quasars detected by LAMOST in 2011 (grey),
   including 509 newly discovered and 23 previously known quasars detected by LAMOST.
   The $i$-band magnitude distributions of the SDSS DR7 quasars are shown here for comparison (black),
   the steep gradients at $i \approx 19.1$ and $i \approx 20.2$ are due to the magnitude limits of
   the SDSS selection algorithms for the low- and high-redshift quasars.
   The magnitude bin size is set to 0.1. The y-axis is displayed in quasar number density.
   \label{fig:mag}}
\end{figure}

\begin{figure}
\includegraphics[width=13.5cm,angle=0]{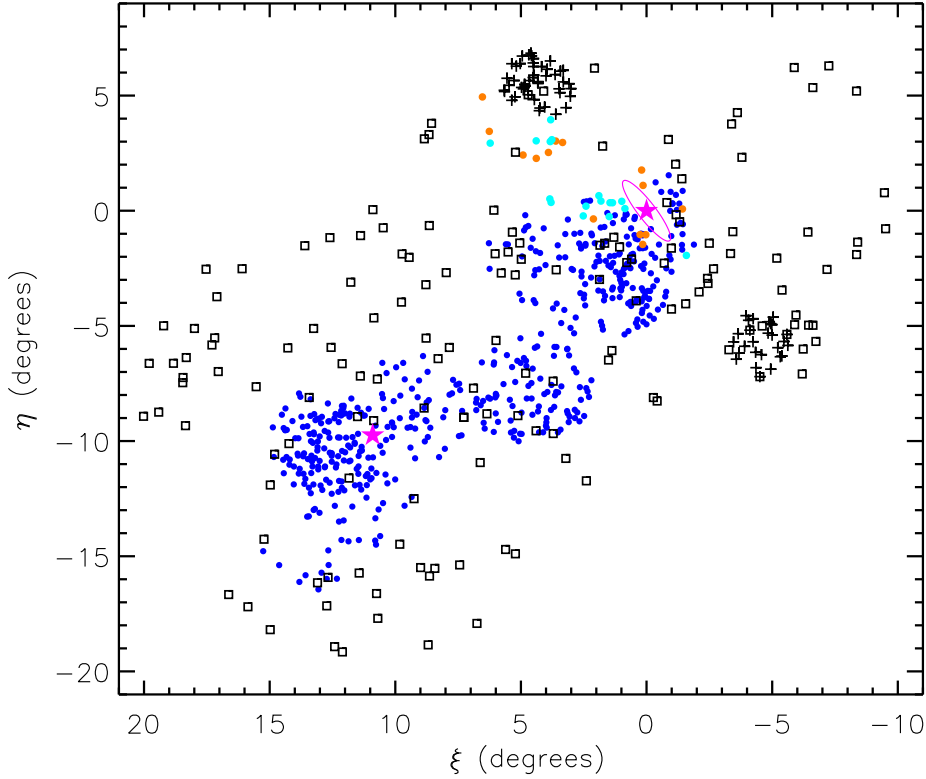}
\caption{Spatial distributions of background quasars in the vicinity of
M\,31 and M\,33. Orange, cyan and blue filled circles represent quasars
identified in the LAMOST 2009, 2010
and 2011 datasets, respectively. Crosses and open squares represent SDSS
quasars and previously known quasars with redshifts archived in the NED, respectively.
The magenta stars mark the central positions of M\,31 and M\,33, while the magenta ellipse
represents the optical disk of M\,31 of radius $R_{25} = 95\farcm3$. 
\label{fig:spatial}}
\end{figure}

The sample has a magnitude limit of $i=20.0$.  Amongst the 526 newly discovered
quasars, the brightest, J010415.77+402243.9, has an $i$ magnitude 15.45 with
redshift of 0.193, and was previously unknown (its spectrum is shown in
Fig.\,\ref{spectra}).  There are 5/20/75 new quasars brighter than
17.0/17.5/18.0 magnitude in $i$-band, respectively. For comparison, amongst the
quasars discovered in the three SDSS spectroscopic plates (Adelman-McCarthy et
al. 2006, 2007) and in the LAMOST 2009 commissioning observations, the
corresponding numbers are 1/1/12. Within a 10\,deg.\ radius of M\,31 and of M\,33,
there are 155 previously known quasars with redshift estimates reported in
the NED archive (see also Fig.\,\ref{fig:spatial} ). Many of them have multiple
photometric measurements taken in a variety of filters, and 151 of them fall in
the area covered by the SDSS or XSTPS surveys.  Using a 3\arcsec\
cross-correlation radius, we searched their counterparts in the SDSS and XSTPS
sources with $i$ magnitudes brighter than 21.0 and 20.0, respectively. We find
116 of them have $i$ band counterparts, amongst them there are 20/41/52 quasars
brighter than 17.0/17.5/18.0 magnitudes, respectively.  These bright quasars
will serve as good candidates for absorption-line spectroscopic studies of the
ISM of M\,31, M\,33 and of their related substructures. Fig.\,\ref{fig:mag}
shows the magnitude distributions of the 532 quasars detected in the 2011
datasets, including 509 newly discovered quasars and 23 previously
known sources which were re-observed by LAMOST. The magnitude has a
bin size of 0.1, the y-axis is in
quasar number density. In the magnitude distributions, as well as the following
redshift distributions, only quasars detected in $\sim$ 135 sq.\ deg in 2011 are
included, the quasars detected in 2010 are not considered. The magnitude
distributions of the SDSS DR7 quasars (Schneider et al.\ 2010) are also shown
for comparison. We can see that a large number of quasars with $i$ magnitude
fainter than 18.0 are left out by LAMOST due to its relatively low performance
during the commissioning and pilot survey phases.


\begin{figure} 
\includegraphics[width=13.5cm, angle=0]{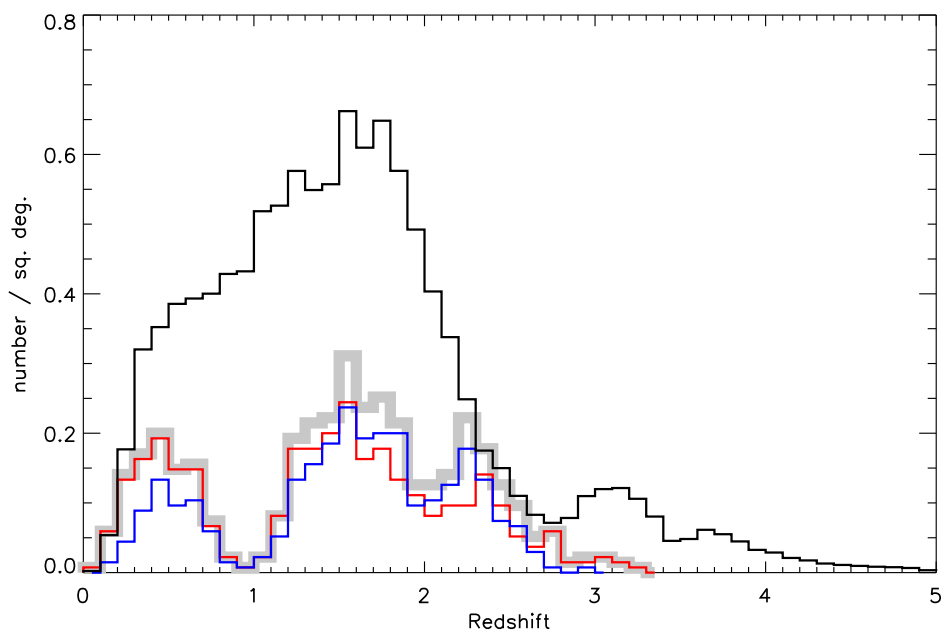}
\caption{Redshift distributions of the 532 LAMOST detected quasars in 2011 (thick grey).
   The blue and red lines represent
   redshift distributions of SDSS low-redshift and optical/IR selected quasars, respectively.
   The redshift distributions of the SDSS DR7 quasars are also shown here for comparison (black),
   the low number density at $z \sim 2.7$ is caused by the degeneracy of the SDSS color of stars
   and colors of quasars near this redshift.
   The redshift bin size is set to 0.1. The y-axis is displayed in quasar number density.
   \label{fig:redshift}}
\end{figure}

\begin{figure} 
\includegraphics[width=13.5cm, angle=0]{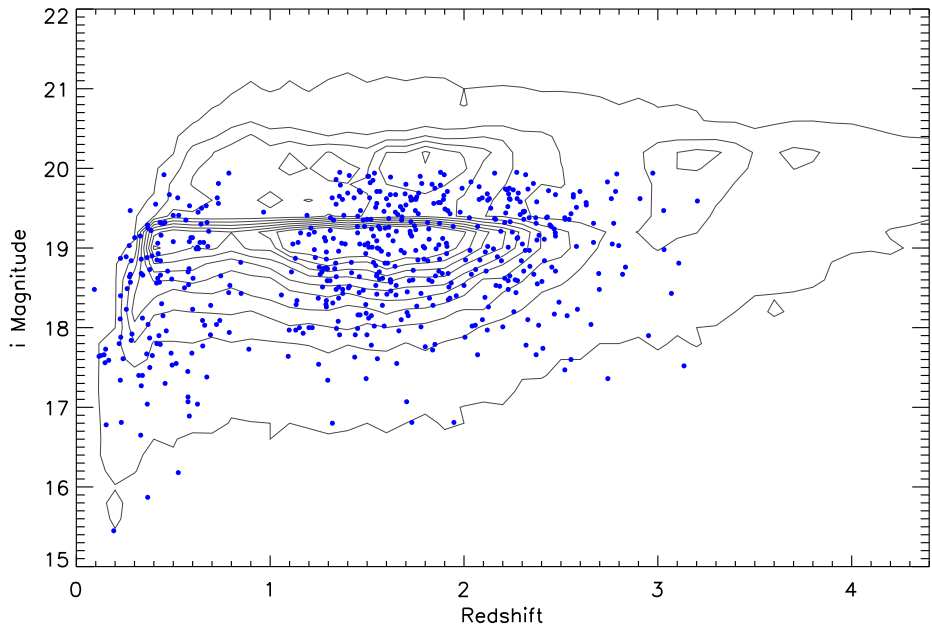}
\caption{Redshift versus $i$-band magnitude distributions of the 532 LAMOST detected
   quasars in 2011 (blue points).
   The contours represent the redshift versus $i$-band magnitude distributions of the SDSS DR7
   quasars. The steep gradients at $i \approx 19.1$ and $i \approx 20.2$ are due to the magnitude limits of
   the SDSS selection algorithms for the low- and high-redshift quasars; the low number density at $z \sim 2.7$ is
   caused by the degeneracy of the SDSS color of stars and colors of quasars near this redshift.
   \label{fig:mag-redshift}}
\end{figure}

Fig.\,\ref{fig:redshift} shows the redshift distributions of the 532 quasars
detected by LAMOST in 2011.  The redshift has a bin size of 0.1, the y-axis is
in quasar number density, as in Fig.\,\ref{fig:mag}.  The lowest
redshift is found for J003842.80+384551.9, at $z=0.093$, whilst the
highest redshift object is J011118.08+325833.8, at $z=3.204$. 
For the optical selection criteria
adopted here, low-redshift quasars selected as outliers of stellar loci on the
SDSS color-color diagrams are mainly $z \leq 2.2$ quasars (Fan 1999; Richards
et al.\ 2001), whereas the optical/IR algorithm is capable of finding quasars of
redshifts up to $z < 3.5$ (Wu et al.\ 2012).  We notice that there is no
significant difference in redshift distribution between the optical and
optical/IR selected quasars, as illustrated by the blue and red lines in
Fig.\,\ref{fig:redshift},  respectively.  For comparison, we also plot the
redshift distributions of SDSS DR7 quasars.  Except for the relatively
lower number-density in redshift space, we can also clearly see that there are two troughs
in the redshift distributions of LAMOST detected quasars (see
Fig.\,\ref{fig:redshift}), one near redshift $\sim $ 0.9 and another near
$\sim$ 2.1. These are entirely due to the selection effects.
For redshift $z < 0.9$, the Mg~{\sc ii} $\lambda$2800, H$\delta$, H$\gamma$,
H$\beta$ and H$\alpha$, as well as the forbidden lines  [O~{\sc iii}]
$\lambda$4959 and $\lambda$5007 all fall within the LAMOST wavelength coverage.
For redshifts $1.1 < z < 2.0$, the C~{\sc iv} $\lambda$1549, C~{\sc iii}]
$\lambda$1908, and Mg~{\sc ii} $\lambda$2800 lines are redshifted into the
LAMOST wavelength range. For redshifts $z > 2.1$, Ly$\alpha$ and N~{\sc v},
Si~{\sc iv} and C~{\sc iv} resonant lines as well as the C~{\sc iii}] $\lambda$1908 
line become visible.
For quasars near the $z \sim$ 0.9 trough, only the Mg~{\sc ii}
$\lambda$2800 line is
easily detectable, other emission lines such as H$\delta$, H$\gamma$ and
H$\beta$ are all redshifted to wavelengths longer than 7000 {\rm \AA}, where
the spectra are seriously contaminated by telluric emission lines. The
difficulty remains until the redshift reaches $\rm z \sim$ 1.1 when the C~{\sc
iii}] $\lambda$1908 becomes visible.  
At redshifts $\sim$ 2.1, although there
are three emission lines, the C~{\sc iv} $\lambda$1549, C~{\sc iii}] $\lambda$1908, 
and Mg~{\sc ii} $\lambda$2800 lines fall in the optical, the  C~{\sc
iii}] $\lambda$1908 line is unfortunately redshifted to a wavelength
of 5900 ${\rm
\AA}$, the cross-over wavelength of the dichroic  where the instrument has a
very low throughput, and the Mg~{\sc ii} $\lambda$2800 is redshifted to a
wavelength $\sim $ 8700 ${\rm \AA}$ where sky emission lines dominate the
region.  For the 213 `probable' quasar candidates detected in the LAMOST 2010
and 2011 datasets, either only one emission line had been detected or the
spectrum was of marginal S/N.  Since we were unable to discriminate the
C~{\sc iv} $\lambda$1549 and Mg~{\sc ii} $\lambda$2800 lines, no redshifts were
provided.  
Fig.\,\ref{fig:mag-redshift} shows the distributions of quasars in
redshift-magnitude space, we can see that not only the selection effects in
redshift space, but also the low number density of quasars for $i$-band fainter
than 18.0, indicate that the performance of LAMOST and 2D pipeline need to be
carefully characterized and optimized.

Fig.\,\ref{fig:spatial} plots the spatial distribution in the $\xi$--$\eta$
plane of all known background quasars found in the vicinity of M\,31 and M\,33,
including the 526 newly identified with the LAMOST which are reported in this paper, 14
quasars discovered by the LAMOST based on earlier commissioning observations
(Paper I), 75 SDSS quasars (Adelman-McCarthy et al.\ 2006, 2007) and 155
previously known quasars with redshifts listed in the NED archive.  Here $\xi$
and $\eta$ are respectively Right Ascension and Declination offsets relative to
the optical center of M\,31 (Huchra, Brodie \& Kent 1991). In
Fig.\,\ref{fig:spatial}, the magenta stars denote the central positions of
M\,31 and M\,33, respectively. The magenta ellipse represents the optical disk
of M\,31 with an optical radius $R_{25} = 95\farcm3$ (de Vaucouleurs et
al.\ 1991), an inclination angle $i = 77$\,deg.\ and a position angle $PA$ =
35\,deg.\ (Walterbos \& Kennicutt 1987).  The quasars are distributed around
M\,31 and M\,33, along the Giant Stellar Stream (McConnachie et al.\ 2009), as
well as the extended halo of M\,31.  Amongst them, there are 93 quasars within
2.5$^\circ$ of M\,31 (about 34\,kpc assuming a distance of 785 kpc to M\,31;
McConnachie et al.\ 2005), 73 of them are newly discovered, 7 are from Paper I
and 13 are previously known quasars listed in the NED. Several quasars fall
very close to the optical disk of M\,31.  Tens of quasars behind the Giant
Stellar Stream have been found, and the number of known quasars behind M\,31's
extended halo and its associated substructures increases by a substantial
amount.  The much-enlarged number of known quasars in the vicinity of M\,31 and
M\,33 should provide an invaluable sample for future PM and ISM/IGM studies of M\,31,
M\,33 and the Local Group.

No quasars have been identified near the central regions of M\,31 and M\,33 due
to crowding in the field.  Outside those regions, the spatial distributions of
identified quasars is not uniform in the 11 fields observed in 2011,
one covering M\,33 yielded 147 quasars, giving the highest quasar number
density of $\sim$7.5/sq. deg. On the other hand, a plate near the bottom-left corner
of Fig.\,\ref{fig:spatial} yielded only 34 quasars.  The quasar number density
identified by the LAMOST is lower than that of SDSS for low-redshift quasars
with point-like morphology, approximately 10 deg${^{\rm -2}}$ (Schneider et al.\
2010).  The low yields were mainly due to insufficient S/N, either because of
poor weather conditions or lower than expected performance of the LAMOST.
Since the current sample of quasars were identified using the commissioning and
pilot survey data, the quasar selection algorithm and telescope performance was
still undergoing development and testing, hence the sample is not homogeneous and is
not intended for statistical analysis.

\begin{acknowledgements}

This work is partially supported by the Young Researcher Grant of National Astronomical
Observatories, Chinese Academy of Sciences, and China Postdoctoral Science
Foundation. We would like to thank Prof. Xuebing Wu and Xiaohui Fan for their careful 
reading and providing valuable comments and suggestions of this
paper. Wu would also like to thank Dr. Stephen Justham for improving the writing of the paper.

The Guoshoujing Telescope (the Large Sky Area Multi-Object Fiber Spectroscopic
Telescope; LAMOST) is a National Major Scientific Project built by the Chinese
Academy of Sciences. Funding for the project has been provided by the National
Development and Reform Commission. The LAMOST is operated and managed by the
National Astronomical Observatories, Chinese Academy of Sciences.

Funding for the SDSS and SDSS-II has been provided by the Alfred P. Sloan
Foundation, the Participating Institutions, the National Science Foundation,
the U.S. Department of Energy, the National Aeronautics and Space
Administration, the Japanese Monbukagakusho, the Max Planck Society, and the
Higher Education Funding Council for England.  The SDSS Web Site is
http://www.sdss.org/.

This publication makes use of data products from the Wide-field Infrared Survey
Explorer, which is a joint project of the University of California, Los
Angeles, and the Jet Propulsion Laboratory/California Institute of Technology,
funded by the National Aeronautics and Space Administration.

This research has made use of the NASA/IPAC Extragalactic Database (NED) which
is operated by the Jet Propulsion Laboratory, California Institute of
Technology, under contract with the National Aeronautics and Space
Administration.

\end{acknowledgements}



\begin{thebibliography}{99}
\small \setlength{\itemindent}{-3mm} \setlength{\itemsep}{-0.5mm}
\setlength{\baselineskip}{4.5mm}


\bibitem[Adelman-McCarthy et al.(2006)]{2006ApJS..162...38A} Adelman-McCarthy, J.~K., et al.\ 2006, \apjs, 162, 38
\bibitem[Adelman-McCarthy et al.(2007)]{2007ApJS..172..634A} Adelman-McCarthy, J.~K., et al.\ 2007, \apjs, 172, 634
\bibitem[Aihara et al.(2011)]{2011ApJS..193...29A} Aihara, H., Allende Prieto, C., An, D., et al.\ 2011, \apjs, 193, 29
\bibitem[Binney \& Tremaine(1987)]{1987gady.book.....B} Binney, J., \& Tremaine, S.\ 1987, Princeton, NJ, Princeton University Press, 1987, 747 p. 605
\bibitem[Brunthaler et al.(2005)]{2005Sci...307.1440B} Brunthaler, A., Reid, M.~J., Falcke, H., Greenhill, L.~J., \& Henkel, C.\ 2005, Science, 307, 1440
\bibitem[Chemin et al.(2009)]{Chemin et al.(2009)} Chemin, L., Carignan, C., \& Foster, T.\ 2009, \apj, 705, 1395
\bibitem[Crampton et al.(1997)]{1997AJ....114.2353C} Crampton, D., Gussie, G., Cowley, A.~P., \& Schmidtke, P.~C.\ 1997, \aj, 114, 2353
\bibitem[Cui et al.(2004)]{2004SPIE.5489..974C} Cui, X., et al.\ 2004, \procspie, 5489, 974
\bibitem[Cui et al.(2010)]{2010SPIE.7733E...7C} Cui, X., Su, D.-Q., Wang, Y.-N., et al.\ 2010, \procspie, 7733, 7
\bibitem[Cui et al.(2012)]{2012RAA....12.1197C} Cui, X.-Q., Zhao, Y.-H., Chu, Y.-Q., et al.\ 2012, Research in Astronomy and Astrophysics, 12, 1197
\bibitem[Darling(2011)]{2011ApJ...732L...2D} Darling, J.\ 2011, \apjl, 732, L2
\bibitem[de Vaucouleurs et al. (1991)]{de Vaucouleurs et al. (1991)} de Vaucouleurs, G., de Vaucouleurs, A., Corwin, H.~G., Jr., Buta, R.~J., Paturel, G., \& Fouqu\'{e}, P.\ 1991, Third Reference Catalogue of Bright Galaxies (New York: Springer)
\bibitem[Fan(1999)]{1999AJ....117.2528F} Fan, X.\ 1999, \aj, 117, 2528
\bibitem[Gilbert et al. 2009]{Gilbert et al. 2009} Gilbert, K.~M., Guhathakurta, P., Kollipara, P., Beaton, R.~L., Geha, M.~C., Kalirai, J.~S., Kirby, E.~N., Majewski, S.~R., \& Patterson, R.~J.\ 2009, \apj, 705, 1275
\bibitem[Huchra et al.(1991)]{Huchra et al.(1991)} Huchra, J. P., Brodie, J. P., \& Kent, S. M.\ 1991, \apj, 370, 495
\bibitem[Huo et al.(2010)]{2010RAA....10..612H} Huo, Z.-Y., Liu, X.-W., Yuan, H.-B., et al.\ 2010, Research in Astronomy and Astrophysics, 10, 612
\bibitem[Ibata et al. (2005)]{Ibata et al. (2005)} Ibata, R., Chapman, S., Ferguson, A.~M.~N., Lewis, G.~F., Irwin, M.~J., \& Tanvir, N.\ 2005, \apj, 634, 287
\bibitem[Ibata et al. (2007)]{Ibata et al. (2007)} Ibata, R., Martin, N.~F., Irwin, M.~J., Chapman, S., Ferguson, A.~M.~N., Lewis, G.~F., \& McConnachie, A.~W.\ 2007, \apj, 671, 1591
\bibitem[Jester et al.(2005)]{2005AJ....130..873J} Jester, S., Schneider, D.~P., Richards, G.~T., et al.\ 2005, \aj, 130, 873
\bibitem[Kim et al.(2012)]{2012ApJ...747..107K} Kim, D.-W., Protopapas, P., Trichas, M., et al.\ 2012, \apj, 747, 107
\bibitem[Koz{\l}owski et al.(2012)]{2012ApJ...746...27K} Koz{\l}owski, S., Kochanek, C.~S., Jacyszyn, A.~M., et al.\ 2012, \apj, 746, 27
\bibitem[Loeb et al.(2005)]{2005ApJ...633..894L} Loeb, A., Reid, M.~J., Brunthaler, A., \& Falcke, H.\ 2005, \apj, 633, 894
\bibitem[Luo et al.(2012)]{2012RAA....12.1243L} Luo, A.-L., Zhang, H.-T.,  Zhao, Y.-H., et al.\ 2012, Research in Astronomy and Astrophysics, 12, 1243
\bibitem[Luo et al.(2004)]{2004SPIE.5496..756L} Luo, A.-L., Zhang, Y.-X., \& Zhao, Y.-H.\ 2004, \procspie, 5496, 756
\bibitem[Massey et al.(2006)]{2006AJ....131.2478M} Massey, P., Olsen, K.~A.~G., Hodge, P.~W., et al.\ 2006, \aj, 131, 2478
\bibitem[McConnachie et al.(2005)]{2005MNRAS.356..979M} McConnachie, A.~W., Irwin, M.~J., Ferguson, A.~M.~N., Ibata, R.~A., Lewis, G.~F., \& Tanvir, N.\ 2005, \mnras, 356, 979
\bibitem[McConnachie et al.(2009)]{McConnachie et al.(2009)} McConnachie, A.~W., Irwin, M.~J., Ibata, R.~J., et al.\ 2009, Nature, 461, 66
\bibitem[Richards et al.(2001)]{2001AJ....121.2308R} Richards, G.~T., Fan, X., Schneider, D.~P., et al.\ 2001, \aj, 121, 2308
\bibitem[Richards et al. (2002)]{Richards02} Richards, G.~T., et al.\ 2002, \aj, 123, 2945
\bibitem[Savage et al.(2000)]{2000ApJS..129..563S} Savage, B.~D., Wakker, B., Jannuzi, B.~T., et al.\ 2000, \apjs, 129, 563
\bibitem[Schlegel et al.(1998)]{1998ApJ...500..525S} Schlegel, D.~J., Finkbeiner, D.~P., \& Davis, M.\ 1998, \apj, 500, 525
\bibitem[Schneider et al.(1993)]{1993ApJS...87...45S} Schneider, D.~P., Hartig, G.~F., Jannuzi, B.~T., et al.\ 1993, \apjs, 87, 45
\bibitem[Schneider et al. (2010)]{2010ApJ...139...2360} Schneider, D.~P., Richards, G.~T., \& Hall, P.~B. et al.\ 2010, \apj, 139, 2360
\bibitem[Sohn et al.(2012)]{2012ApJ...753....7S} Sohn, S.~T., Anderson, J., \& van der Marel, R.~P.\ 2012, \apj, 753, 7
\bibitem[Stern et al.(2012)]{2012ApJ...753...30S} Stern, D., Assef, R.~J., Benford, D.~J., et al.\ 2012, \apj, 753, 30
\bibitem[Su et al.(1998)]{1998SPIE.3352...76S} Su, D.~Q., Cui, X., Wang, Y., \& Yao, Z.\ 1998, \procspie, 3352, 76
\bibitem[Tinney et al.(1997)]{1997MNRAS.285..111T} Tinney, C.~G., Da Costa, G.~S., \& Zinnecker, H.\ 1997, \mnras, 285, 111
\bibitem[Tinney(1999)]{1999MNRAS.303..565T} Tinney, C.~G.\ 1999, \mnras, 303, 565
\bibitem[van der Marel \& Guhathakurta(2008)]{2008ApJ...678..187V} van der Marel, R.~P., \& Guhathakurta, P.\ 2008, \apj, 678, 187
\bibitem[Wang et al.(1996)]{1996ApOpt..35.5155W} Wang, S.-G., Su, D.-Q., Chu, Y.-Q., Cui, X., \& Wang, Y.-N.\ 1996, \ao, 35, 5155
\bibitem[Walterbos and Kennicutt(1987)]{Walterbos and Kennicutt(1987)} Walterbos, R.~A.~M., \& Kennicutt, R.~C., Jr.\ 1987, A\&AS, 69, 311
\bibitem[Wright et al.(2010)]{2010AJ....140.1868W} Wright, E.~L., Eisenhardt, P.~R.~M., Mainzer, A.~K., et al.\ 2010, \aj, 140, 1868
\bibitem[Wu et al.(2012)]{2012AJ....144...49W} Wu, X.-B., Hao, G., Jia, Z., Zhang, Y., \& Peng, N.\ 2012, \aj, 144, 49
\bibitem[Xing et al.(1998)]{1998SPIE.3352..839X} Xing, X., Zhai, C., Du, H., Li, W., Hu, H., Wang, R., \& Shi, D.\ 1998, \procspie, 3352, 839
\bibitem[Yan et al.(2013)]{2013AJ....145...55Y} Yan, L., Donoso, E., Tsai, C.-W., et al.\ 2013, \aj, 145, 55
\bibitem[York et al. 2000]{York2000} York, D.~G., et al.\ 2000, \aj, 120, 1579
\bibitem[Zhao et al.(2012)]{2012RAA....12..723Z} Zhao, G., Zhao, Y.-H., Chu, Y.-Q., Jing, Y.-P., \& Deng, L.-C.\ 2012, Research in Astronomy and Astrophysics, 12, 723
\bibitem[Zhao(2000)]{2000SPIE.4010..290Z} Zhao, Y.\ 2000, \procspie, 4010, 290
\bibitem[Zhu et al.(2006)]{2006SPIE.6269E..20Z} Zhu, Y., Hu, Z., Zhang, Q., Wang, L., \& Wang, J.\ 2006, \procspie, 6269, 20

\end{thebibliography}
\end{document}